%% file: noise-satis.tex
\documentclass[10pt,conference]{ieeeconf}
\usepackage{amsmath,epsfig,alltt}
\usepackage{amssymb}
\usepackage{setspace}
\usepackage{rotating}
\usepackage{subfigure}
\usepackage{algorithmic}
\usepackage{algorithm}

\usepackage{multirow}
\usepackage{mathptm}
\usepackage{spk}
\usepackage{color}

\newtheorem{theorem}{Theorem}[section]

\newtheorem{definition}{Definition}

\begin{document}

\title{Boolean Satisfiability using Noise Based Logic}
\author{
Pey-Chang Kent Lin, Ayan Mandal, Sunil P. Khatri\\
Department of ECE, Texas A\&M University, College Station TX 77843\\
}
\date{}
\maketitle

\thispagestyle{empty}
\begin{spacing}{0.95}
\input{abstract}
\input{intro}

\input{previous}

\input{approach}
\input{experiment}

\input{discussion}

\input{conclusion}

\bibliography{references} 
\bibliographystyle{ieeetr}
\end{spacing}
\end{document}

%% file: abstract.tex
\begin{abstract}

In this paper, we present a novel algorithm to solve the Boolean
Satisfiability (SAT) problem, using noise-based logic (NBL). Contrary to
what the name may suggest, NBL is not a random/fuzzy logic system. In fact,
it is a completely deterministic logic system. A key property of NBL is that it 
allows us to apply a superposition of many input vectors to a SAT instance at
the same time, circumventing a key restriction and assumption in the
traditional approach to solving SAT. 
By exploiting the superposition property of
NBL, our NBL-based SAT algorithm can determine whether an instance is SAT or not in 
a single operation. A satisfying solution can be found by iteratively performing SAT check operations
up to $n$ times, where $n$ is the number of variables in the SAT instance. 
Although this paper does not focus on the realization of an NBL-based SAT
engine, such an engine can be conceived using analog circuits (wide-band amplifiers, adders and multipliers),
FPGAs or ASICs. Additionally, we also discus scalability of our approach, which can apply 
to NBL in general.
The NBL-based SAT engine described in this paper
has been simulated in software for validation purposes.



\end{abstract}

%% file: intro.tex
\section{Introduction}
\label{sec:intro}

Boolean Satisfiability (SAT)~\cite{cook} is a core NP-complete problem which
has been studied extensively. Given a set
$V$ of variables ($n$ in all), and a collection $C$ of Conjunctive Normal
Form (CNF) clauses  over $V$ ($m$ in all), the SAT problem consists of determining if
there is a satisfying truth assignment for $C$, and returning this truth
assignment. The CNF expression $C$ is referred to as a {\em SAT
  instance}. If no satisfying assignment exists, $C$ is referred to as an {\em
  unsatisfiable instance}.   

The applicability of SAT to several problem domains
such as logic synthesis, formal verification, circuit testing, pattern 
recognition and others~\cite{gu} has resulted in much effort devoted to
devising   efficient heuristics to solve SAT.
Some of the more well-known complete approaches for SAT
include~\cite{grasp, zchaff, berkmin, circus} and~\cite{Minisat}.
In addition, several incomplete or stochastic heuristics have been
developed as well. A partial list of these are~\cite{walksat, gsat,
  Shang98adiscrete, sp1, sp2}. The complete approaches seek to find a
satisfying solution (or to prove that none exists) by heuristically
assigning a logic variable of the problem to ''1'' or ''0''. By analyzing
the consequences of such an assignment, a new variable is assigned, or a
previously assigned variable is backtracked upon. This is continued until
$C$ is satisfied, or the search space is exhausted (in which case $C$ is
proven to be unsatisfiable)

Recently, it was shown that noise can be used to realize logic
circuits~\cite{lkthn1, lkthn2, sethu}. We refer to this logic scheme as
Noise-based Logic (NBL) in the sequel. In NBL, a plurality of
pairwise uncorrelated noise sources (referred to as {\em noise bits}) are
utilized. Each such noise source has zero mean, and all the sources have
the same RMS value (assumed to be zero). It is important to point out that
NBL is a {\bf 
  deterministic logic scheme}, and {\em not} fuzzy or probabilistic in
nature. NBL can be utilized to realize multi-valued logic as
well~\cite{sethu, Bollapalli2010}.

The orthogonality property of  the noise bits yields some powerful properties:
\begin{itemize} 
\item Starting with $2n$ pairwise orthogonal basis noise sources, we can
  create a noise hyperspace of cardinality $2^n$, by appropriately multiplying 
  these 
noise sources~\cite{sethu}. On a single wire, the additive superposition of any subset
of this hyperspace can be transmitted, and this yields a total of 
$2^{2^n}$ possible logic values that can be transmitted on the wire. In
effect the wire behaves like $2^n$ wires carrying binary valued signals.

\item In addition, we can apply all possible inputs to an $n$ input NBL circuit {\em
    simultaneously}. Consider a combinational circuit with $n$ inputs $x_1,
    x_2, \cdots x_n$. For each input $x_i$, let us assume we have a noise
    source (noise bit) $N_{\overline{x_i}}$ to represent the $\overline{x_i}$ literal, and a noise
    source $N_{x_i}$ to represent the $x_i$ literal. Hence, for $1 \le i \le
    n$, we may apply the input ($N_{x_i} + N_{\overline{x_i}}$) to the $i^{th}$ 
    input of the circuit. This in effect means that we applied {\em all} $2^n$
    inputs to the circuit {\em simultaneously}. We will see how this ends
    up being very important in Section~\ref{sec:approach}.

\end{itemize}

In this paper, we present an approach to solve the SAT problem utilizing
NBL. The resulting approach can provide a SAT/UNSAT decision in a single operation, and can provide a satisfying input vector in a number of such operations which is linear in $n$. This 
is possible because NBL allows us to apply all $2^n$
inputs to the circuit {\em simultaneously}.  Although no NBL circuits
exist today, realizing the NBL-SAT solution approach of our paper
would require widely studied, and ubiquitously available circuit components
such as wideband amplifiers, analog 
adders and analog multipliers and low-pass filters.
NBL-SAT may be implemented on FPGAs or ASICs as well.
We hope that the result
of this paper will encourage development of NBL circuits. 



Before we list the contributions of this paper, we would like to reiterate that the NBL used in this paper is not
probabilistic or fuzzy. Rather it 
is completely deterministic logic scheme. So the claims made in this
paper are not probabilistic or fuzzy, but are completely deterministic.

The key contributions of this paper are:
 \begin{itemize}
 \item We show how NBL can solve the SAT problem. The resulting algorithm
   can determine if a problem is SAT or UNSAT in one operation, and can provide
   a satisfying assignment in $n$ operations, where $n$ is number of variables
   in the SAT problem.
 \item Although the focus of this paper is not to provide concrete
   realizations of the NBL based SAT algorithm, we show that such
   realizations are imminently realizable with existing technology
   \begin{itemize}
     \item A hardware based NBL-SAT solver requires commonly available
       hardware components such as wide-band amplifiers, analog
       multipliers, analog adders, and low-pass filters, or even FPGAs or ASICs.
     \item A software based NBL-SAT solver can be envisioned. Initial
       proof-of-correctness simulations of our algorithm were done on a
       MATLAB based realization of our algorithm. 
     \item Alternative realizations of NBL-SAT can be envisioned using
       sinusoidal signals~\cite{Bollapalli2010} or Random Telegraph Waves~\cite{Kish2010}.
   \end{itemize} 
 \end{itemize}

The remainder of this paper is organized as
follows. Section~\ref{sec:previous} 
discusses some previous work in this area. In Section~\ref{sec:approach} we
describe our approach to solving the SAT problem using
NBL. Section~\ref{sec:experiment} presents experimental results from a
MATLAB based simulation which validates our approach.
Section~\ref{sec:discussion} discusses possible realizations of NBL-SAT,
while conclusions are discussed in Section~\ref{sec:conclusion}.

%% file: previous.tex
\section{Previous Work}
\label{sec:previous}

The idea of noise based logic was recently developed, and initially
described in~\cite{lkthn1}. In~\cite{lkthn2}, the concept of NBL was
extended to multi-valued signals as well, and it was shown that sinusoidal
tones could be used instead of uncorrelated noise signals, as the
information carriers. The idea of using an additive superposition of noise bits to
generate a noise-based hyperspace idea was presented in~\cite{sethu}. 
Starting with $2n$ basis noise sources, it was shown how one could
construct a hyperspace of $2^n$ noise sources using a linear number of additions and multiplications.
By an additive superposition
of any subset of this hyperspace, it was shown how a single wire could
carry as many as $2^{2^n}$ symbols in it, effectively accomplishing the
task of  $2^n$ binary-valued wires.

Several derivative papers~\cite{Bezrukov2009, Gingl2010, Kish2010} of these works developed the
idea of noise based logic further, using pulse based~\cite{Bezrukov2009, Gingl2010} or Random
Telegraph Wave (RTW)~\cite{Kish2010}  based signals. A VLSI implementation of
NBL (using sinusoidal signals as information carriers) was presented
in~\cite{Bollapalli2010}. In this paper, the specialized and restricted version of
NBL used sinusoids. In particular the logic 1 and logic 0 signals were
chosen to be anti-correlated, in order to mimic binary logic and demonstrate
the viability of the approach. It was shown that with existing MOSFETs, one
can realize gates using sinusoidal logic. To the best of the authors'
knowledge, there has been no effort to date, to realize true NBL gates or
circuits.  

Just like NBL, quantum computers have the capability of applying a
superposition of input values to a quantum circuit. In the past, there has
been work in the realm of quantum computing~\cite{ohya2005, ohya2000, ohya2003}
focusing on solving SAT on quantum computers. There are some precise
differences between these papers and the NBL based SAT engine described in
this paper:

\begin{itemize}
\item In contrast to our approach,~\cite{ohya2005, ohya2000, ohya2003} only solve the
  problem of determining whether a SAT instance is satisfiable or
  unsatisfiable. Our approach, over and above that of~\cite{ohya2005,
    ohya2000, ohya2003}, provides an {\em algorithm to determine the
  satisfying assignment (if one exists) using a linear number of NBL-SAT checks}.


\item Also, our NBL based SAT algorithm is realizable using existing
  ubiquitous circuit components (such as wide-band amplifiers, analog adders, analog
  multipliers and filters).  In contrast, the field of quantum computing is
  extremely young, without the ability to realize even medium sized quantum
  circuits, severely hampering the applicability of the quantum SAT
  algorithm of~\cite{ohya2005, ohya2000, ohya2003}.
\end{itemize}

%% file: approach.tex
\section{Our Approach}
\label{sec:approach}

Before discussing our NBL-SAT algorithm, we first provide definitions related to topics of Boolean Satisfiability, NP-completeness, and Noise-based Logic.

\subsection{Definitions}

\begin{definition}
A {\bf literal} or a {\bf literal function} is a binary variable $x$ or its
negation $\overline{x}$.
\end{definition}

\begin{definition}
A {\bf cube} is a conjunction (AND) of one or more literal functions, i.e. $x_1x_2\overline{x_3}$.
\end{definition}

\begin{definition}
A {\bf clause} is a disjunction (OR) of one or more literal functions, i.e. $(x_1+x_2)$.
\end{definition}

\begin{definition} A {\bf Conjunctive Normal Form (CNF)} formula
consists of a conjunction of $m$ clauses $c_1, c_2 \ldots c_m$. Each
clause $c_i$ consists of the disjunction of $k_i$ literals.
\end{definition}

\begin{definition} A CNF formula is said to be {\bf satisfied} if each of the $m$ clauses of the CNF formula simultaneously evaluate to {\em true}. 
\end{definition}

A CNF formula is also referred to as a logical product of sums. Thus, to
{\em satisfy} the
CNF formula, at least one literal in each clause must evaluate to {\em true}.

\begin{definition} {\bf Boolean Satisfiability (SAT):} Given a Boolean formula $S$
on a set of binary variables $X = \{ x_1, x_2, x_3 \cdots x_n\}$, expressed
in Conjunctive Normal Form (CNF), the objective of SAT is to identify an
assignment of the binary variables in $X$ that satisfies $S$. If no such
assignment exists, this should be indicated. 
\end{definition}

For example, consider the formula $S(x_1,x_2,x_3) =
(x_1+\overline{x_2})\cdot(x_1+x_2+x_3)$. This formula consists of 3 variables, 2 clauses,
and 4 literals. This particular formula is satisfiable, and a satisfying
assignment is $<x_1,x_2,x_3>\,=\,<0,0,1>$, which can be expressed as
the satisfying cube $\overline{x_1}\,\overline{x_2}\,x_3$. The CNF expression $S$ is
often referred to as a {\em SAT instance} in the literature. 

SAT is one of the most well known NP-complete problems. As such, therefore, 
there are no known polynomial time algorithms to solve SAT.
Note that the definition of NP-completeness is premised on the assumption
that a Universal Turing Machine (UTM) is used to perform operations to solve the decision problem $C$.  
In this paper, we sidestep this particular assumption. In particular, using
NBL to solve SAT, we are able to apply a superposition of {\em all} inputs (candidate solutions)
to the problem instance. This superposition property allows us to verify
all solutions simultaneously to determine if the problem is satisfiable 
(or not) in a single operation, and if satisfiable, to provide a satisfying
solution in a number of operations that is linear in $n$.

The remainder of this sub-section presents some definitions pertaining to
Noise-based Logic (NBL).

\begin{definition} {\bf Independent Noise Processes:} Consider two
    noise processes $V_i(t)$ and $V_j(t)$. These noise processes are
    independent iff the  correlation operator $\langle \rangle$ applied
    to $V_i(t)$ and  $V_j(t)$ yields 

 $\langle V_i(t) V_j(t) \rangle = \delta_{i,j}$

where $\delta_{i,j}$ is the Kronecker symbol ($\delta_{i,j} = 1$ when $i =
j$, and $\delta_{i,j} = 0$ otherwise. 
\end{definition}

\begin{definition} {\bf Basis (Reference) Noise Processes (Bits):}
    Consider $M$ noise processes $V_1(t), V_2(t), \cdots, V_M(t)$. If
    these processes are pairwise independent, then $V_1(t), V_2(t), \cdots,
    V_M(t)$ are referred to as basis (reference) noise processes (bits).
\end{definition}

For convenience, we assume that all the noise processes in the sequel have
a zero mean, and a zero RMS value.

Consider two orthogonal basis noise bits $V_i(t)$ and $V_j(t)$ ($i \ne
j$). The product $Z_{i,j}(t) = V_i(t) \cdot V_j(t)$ of two orthogonal basis
noise bits is  orthogonal to $V_k(t)$ ($k = 1, 2, \cdots, M$). This
property was used~\cite{sethu} to realize a {\bf logic hyperspace}. In
other words, 

$\langle Z_{i,j}(t), V_k(t) \rangle$ = 0

\begin{definition} {\bf Noise-based Logic Hyperspace:} Using $2M$ basis noise
bits $V^0_1(t), V^1_1(t), \cdots V^0_m(t), V^1_m(t)$, we can compute
a noise hyperspace $\cal{H}$ with dimensionality $2^M$, by multiplying 
these noise bits appropriately, and performing their additive superposition
as follows:
\end{definition}
\begin{eqnarray*}
\cal{H} & = & V^0_1(t) \cdot V^0_2(t) \cdots V^0_{m-1}(t) + \\
        &   & V^0_1(t) \cdot V^0_2(t) \cdots V^1_{m-1}(t) + \\
        &   &  \cdots + V^1_1(t) \cdot V^1_2(t) \cdots V^1_{m-1}(t)
\end{eqnarray*}
{\bf Example 1:} Consider four orthogonal basis noise bits $V^0_1(t), V^1_1(t), V^0_2(t), V^1_2(t)$.
The noise-based logic hyperspace consists of four hyperspace elements
$V^0_1(t) \cdot V^0_2(t)$,
$V^0_1(t) \cdot V^1_2(t)$,
$V^1_1(t) \cdot V^0_2(t)$,
$V^1_1(t) \cdot V^1_2(t)$.

The power of the noise-based hyperspace is evidenced by the fact that
starting from $2M$ basis noise sources, we can construct a
hyperspace of size $2^M$. Now an additive superposition of any subset of
elements from this hyperspace can be transmitted along a wire. 

In the remainder of this paper, we will refer to noise sources as $N$
instead of $N(t)$. 

\subsection{Generating all $2^n$ Minterms in an NBL Additive Superposition}

Before we describe our NBL based SAT algorithm, we first discuss a means of
constructing the additive superposition of {\em all} input vectors for a
problem~\cite{sethu}. 

Consider a problem on $n$  binary valued variables $x_1, x_2,
\cdots, x_n$. For each variable $x_i$, we define two basis noise sources
$N_{x_i}$ and $N_{\overline{x_i}}$, for the negative and positive literals of $x_i$
respectively. This requires a total of $2n$ basis noise sources. Now, we
can construct the product  

\begin{equation}
T = (N_{x_1} + N_{\overline{x_1}}) \cdot (N_{x_2} + N_{\overline{x_2}}) \cdots (N_{x_n} + N_{\overline{x_n}})
\label{eq:one}
\end{equation}
If $T$ were expanded out, it is easy to see that $T$ is the additive
superposition of $2^n$ products of basis noise sources. Each product
corresponds to a noise-based minterm on the $n$ variable space. 

{\bf Example 3:} Suppose $n = 3$. Then, if $T$ were expanded out, we get 

$T = (N_{x_1} \cdot N_{x_2} \cdot N_{x_3}) +
(N_{x_1} \cdot N_{x_2} \cdot N_{\overline{x_3}}) + 
(N_{x_1} \cdot N_{\overline{x_2}} \cdot N_{x_3}) + 
(N_{x_1} \cdot N_{\overline{x_2}} \cdot N_{\overline{x_3}}) + 
(N_{\overline{x_1}} \cdot N_{x_2} \cdot N_{x_3}) + 
(N_{\overline{x_1}} \cdot N_{x_2} \cdot N_{\overline{x_3}}) + 
(N_{\overline{x_1}} \cdot N_{\overline{x_2}} \cdot N_{x_3}) + 
(N_{\overline{x_1}} \cdot N_{\overline{x_2}} \cdot N_{\overline{x_3}})$

In other words, using Equation~\ref{eq:one}, we are able to generate
the additive superposition of all $2^n$ minterms of the binary space. This
is generated with a linear number of noise sources, and a linear number of
analog adders and multipliers.

An important variation of the above idea is that we can {\em bind} a subset of variables to any literal value in
$T$ above, and generate an additive superposition of the minterms that are
in the cube subspace of the bound variables. In other words, suppose we
bind variables $X = \{ x_i, x_{i+1}, \cdots, x_{i+p-1} \}$ to literals $l_i,
l_{i+1}, \cdots, l_{i+p-1}$ respectively, where $p < n$, then we generate
the additive superposition of all minterms in the cube subspace $l_i \cdot
l_{i+1} \cdot l_{i+2} \cdots l_{i+p-1}$.

{\bf Example 4:} In Example 3, if we bind variable $x_1$ to literal
$\overline{x_1}$, then $T_{\overline{x_1}} = (N_{\overline{x_1}} + 0) \cdot (N_{x_2} + N_{\overline{x_2}}) \cdots (N_{x_n} +
N_{\overline{x_n}})$. If $T_{\overline{x_1}}$ were expanded out, we would get

$T_{\overline{x_1}} = (N_{\overline{x_1}} \cdot N_{x_2} \cdot N_{x_3}) + 
(N_{\overline{x_1}} \cdot N_{x_2} \cdot N_{\overline{x_3}}) + 
(N_{\overline{x_1}} \cdot N_{\overline{x_2}} \cdot N_{x_3}) + 
(N_{\overline{x_1}} \cdot N_{\overline{x_2}} \cdot N_{\overline{x_3}})$

Thus $T_{\overline{x_1}}$ is the additive superposition of all the minterms in the cube
subspace $\overline{x_1}$.

Using the construction of $T$ and $T_v$ subspace above, we now discuss our NBL-SAT
algorithm. 

\subsection{SAT to NBL-SAT Transformation}

In this subsection, we described the process of transforming a SAT decision
problem $S$ into an equivalent NBL formula $S_N$. Consider a decision
problem expressed
as a CNF $S$ with $m$ clauses ($S=c_1 \cdot c_2 \cdots c_m$) on a set of binary variables $X = (x_1, x_2, \cdots, x_n)$. We would like to determine if $S$ is satisfiable, and if so, find a satisfying assignment. 
$S_N$ is comprised of the product of 2 sets of clauses $\tau_N$ and $\Sigma_N$, where $\tau_N$ contains all $2^n$ valid
minterms for the instance $S$, while $\Sigma_N$ contains all satisfying minterms for $S$. These clauses are
discussed in detail in the following.

For each clause $c_j$, 
we create $2n$ independent basis noise sources which are used to represent the positive and negative literal of each
variable $x_1, x_2, \cdots, x_M$. Let $N^j_{x_i}$ be the noise source
corresponding to literal $x_i$ in clause $c_j$, and $N^j_{\overline{x_i}}$ be the noise source
corresponding to literal $\overline{x_i}$ in clause $c_j$.  In total, we create $2mn$ independent 
basis noise sources as there are $m$ clauses, each requiring $2n$ noise sources. Note that the
noise sources are independent across clauses, such that the product of any noise (for any variable $x_p$ and $x_q$)
from clauses $c_j$ and $c_k$
where $j \ne k$ has a zero mean ($N^j_{x_p} \cdot N^k_{x_q} = 0$).

{\bf Construction of $\tau_N$: }First we construct the noise hyperspace $\tau_N$ which contains all $2^n$ valid
minterms to be applied to the SAT instance $\Sigma_N$. The hyperspace $\tau_N$ is constructed 
following Equation~\ref{eq:one}, except the two basis noise sources $N_{x_i}$ and $N_{\overline{x_i}}$
for literals $x_i$ and $\overline{x_i}$ are replaced with the products $N^1_{x_i}N^2_{x_i} \cdots N^m_{x_i}$
and $N^1_{\overline{x_i}}N^2_{\overline{x_i}} \cdots N^m_{\overline{x_i}}$ respectively. These 
products correspond to the product of noise sources for literals $x_i$ and $\overline{x_i}$ 
respectively, used in all clauses for $\Sigma_N$.

\begin{eqnarray}
\tau_N & = & ((N^1_{x_1}N^2_{x_1} \cdots N^m_{x_1}) + (N^1_{\overline{x_1}}N^2_{\overline{x_1}} \cdots N^m_{\overline{x_1}})) \nonumber \\
       &   & \cdot ((N^1_{x_2}N^2_{x_2} \cdots N^m_{x_2}) + (N^1_{\overline{x_2}}N^2_{\overline{x_2}} \cdots N^m_{\overline{x_2}})) \nonumber \\
       &   & \cdots ((N^1_{x_n}N^2_{x_n} \cdots N^m_{x_n}) + (N^1_{\overline{x_n}}N^2_{\overline{x_n}} \cdots N^m_{\overline{x_n}}))
\label{eq:two}
\end{eqnarray}

{\bf Construction of $\Sigma_N$: }Now we construct the NBL-based SAT instance $\Sigma_N$ from the SAT instance $S$ by
replacing the positive literal of variable $v$ in clause $c_j$ by cube subspace $T^j_v$, and
the negative literal of variable $v$ in clause $c_j$ by noise source $T^j_{\overline{v}}$.
By binding the the variable $v$ to the literal value, the cube subspace $T^j_v$ or $T^j_{\overline{v}}$ is an additive superposition of minterms containing the literal value which satisfies clause $c_j$. 

{\bf Example 5:} Consider the CNF formula $S = c_1 \cdot c_2 \cdot c_3 \cdot c_4 = (\overline{x_1}) \cdot (x_2 + x_3) \cdot (x_1 + \overline{x_3}) \cdot (\overline{x_1} + \overline{x_2} + x_3)$. The NBL-SAT instance $\Sigma_N$ is as follows:

$\Sigma_N = (T^1_{\overline{x_1}}) \cdot (T^2_{x_2} + T^2_{x_3}) \cdot (T^3_{x_1} + T^3_{\overline{x_3}}) \cdot (T^4_{\overline{x_1}} + T^4_{\overline{x_2}} + T^4_{x_3})$

When $\Sigma_N$ is expanded out, the noise vectors for minterms from each clause form products with noise
vectors of minterms from all other clauses. A valid satisfying minterm for $\Sigma_N$ would be such
that its final noise product contains a product of noise vectors from all clauses that 
represent the same minterm. All other combination of noise vectors are logically invalid.

Consider a SAT formula $S$ where number of variables and clauses are $n=2$ and $m=2$ respectively.
An example of a valid noise-based minterm is:

$N^1_{x_1}N^2_{x_1}N^1_{\overline{x_2}}N^2_{\overline{x_2}}$

which corresponds to the minterm $x_1\,\overline{x_2}$ of $S$. An example
of an invalid noise-based minterm is:

$N^1_{x_1}N^2_{\overline{x_1}}N^1_{x_2}N^2_{x_2}$

Which corresponds to the term $x_1\,\overline{x_1}\,x_2$ of $S$.

Thus $\Sigma_N$ is the additive superposition of all valid (satisfying) and invalid minterms of the SAT instance.
Since $\tau_N$ only contains all valid minterms as shown in Equation~\ref{eq:two},
the product of $\tau_N \cdot \Sigma_N$ is the additive superposition of the self-correlation of
each of the valid minterms. 
The average value of $\tau_N \cdot \Sigma_N$ is zero if the instance $S$ is unsatisfiable, and positive
if the instance $S$ is satisfiable.

\subsection{Satisfiability Check using NBL-SAT}

Algorithm~\ref{algo:check} describes the procedure for a single operation satisfiability
checking using NBL-SAT. After formulation of NBL-SAT $\Sigma_N$ and hyperspace $\tau_N$, the check for satisfiability is done with an observation of $S_N = \tau_N \cdot \Sigma_N $. If $S_N$ has a zero average, then we conclude $S$ is unsatisfiable. If on the other hand, $S_N$ has a positive average value, then $S$ is satisfiable.

\begin{algorithm}[htb]
 \begin{small}
    \begin{algorithmic}[1]
      \STATE{$NBL-SAT\_check(S_N)$}
      \STATE{$S_N \leftarrow (\tau_N \cdot \Sigma_N)$}
      \IF{$S_N$ output has a zero average}
        \STATE{return($S$ is unsatisfiable)}
      \ELSE
        \STATE{return($S$ is satisfiable)}
      \ENDIF
    \end{algorithmic}
    \caption{Pseudocode of NBL-SAT checker}
    \label{algo:check}
 \end{small}
\end{algorithm}

The key to the single operation SAT check achieved by
this algorithm are the superposition and correlation properties of the noise basis sources. In $\Sigma_N$, each clause $c_j$ contains any number of cube subspaces $T^j_v$. The disjunction of all the $T^j_v$ in clause $c_j$ result in a new noise vector $Z^j$. Thus $Z^j$ is the additive superposition of all noise-based minterms that satisfy clause $c_j$.
Hence $\Sigma_N$ includes the additive superposition of all noise-based minterms that satisfy $S$.
Multiplying $\Sigma_N$ with $\tau_N$ simply yields the additive superposition of the self correlation of the noise-based minterms of $S$.

The output of $\Sigma_N$ is the conjunction (product) of all $Z^j$ noise vectors from the clauses. We recall that the product of two independent noise sources is 0. As $\tau_N$ is the additive superposition of valid minterms for
$S$, then the product $\tau_N \cdot \Sigma_N = 0$ only in the case where $\Sigma_N$ and $\tau_N$ do not share any noise
vectors, and hence, no minterm exists in $\Sigma_N$ that correlates to any of the valid minterms in $\tau_N$.
If $S_N = 0$ or has a zero average, then there is no valid minterm that exists across all clauses
and we conclude $S$ is unsatisfiable (line 4).

However, if $\Sigma_N$ and $\tau_N$ contain common minterm(s), the product of the noise vectors results in a positive average for $S_N$. 
Then if $\tau_N \cdot \Sigma_N$ has a positive DC offset, we can conclude a common satisfying minterm exists across all clauses (line 6).

A demonstration of the algorithm is shown in Examples 6 and 7.

{\bf Example 6:} Consider the CNF formula $S = (x_1 + \overline{x_2}) \cdot (\overline{x_1} + \overline{x_2})$. The NBL-SAT instance is as follows:

$\Sigma_N = (T^1_{x_1} + T^1_{\overline{x_2}}) \cdot (T^2_{\overline{x_1}} + T^2_{\overline{x_2}})$

By expanding $T^j_{x_i}$ to show the minterms:

$\Sigma_N = (N^1_{x_1}N^1_{x_2} + N^1_{x_1}N^1_{\overline{x_2}} + N^1_{\overline{x_1}}N^1_{\overline{x_2}}) \cdot (N^2_{x_1}N^2_{\overline{x_2}} + N^2_{\overline{x_1}}N^2_{x_2} + N^2_{\overline{x_1}}N^2_{\overline{x_2}})$

The minterms $x_1\overline{x_2}, \overline{x_1}\,\overline{x_2}$ exist in all clauses, which will have the noise products $N^1_{x_1}N^2_{x_1}N^1_{\overline{x_2}}N^2_{\overline{x_2}}$, $N^1_{\overline{x_1}}N^2_{\overline{x_1}}N^1_{\overline{x_2}}N^2_{\overline{x_2}}$ respectively in $\Sigma_N$.

The valid minterm hyperspace $\tau_N$ is as follows:

$\tau_N = (N^1_{x_1}N^2_{x_1} + N^1_{\overline{x_1}}N^2_{\overline{x_1}}) \cdot (N^1_{x_2}N^2_{x_2} + N^1_{\overline{x_2}}N^2_{\overline{x_2}})$

$\tau_N = N^1_{x_1}N^2_{x_1}N^1_{x_2}N^2_{x_2} 
+ N^1_{x_1}N^2_{x_1}N^1_{\overline{x_2}}N^2_{\overline{x_2}} 
+ N^1_{\overline{x_1}}N^2_{\overline{x_1}}N^1_{x_2}N^2_{x_2} 
+ N^1_{\overline{x_1}}N^2_{\overline{x_1}}N^1_{\overline{x_2}}N^2_{\overline{x_2}}$

The noise products for the minterms $x_1\,\overline{x_2}$, $\overline{x_1}\,\overline{x_2}$ exist in both $\Sigma_N$ and $\tau_N$. The result $S_N = \tau_N \cdot \Sigma_N$ will be the additive superposition of the self-correlation of these two noise products and $S_N$ will thus have a positive average, concluding this example as satisfiable.

{\bf Example 7:} Consider the CNF formula $S = (x_1) \cdot (\overline{x_1})$. The NBL-SAT instance is as follows:

$\Sigma_N = (T^1_{x_1}) \cdot (T^2_{\overline{x_1}})$

By expanding $T^j_{x_i}$ to show the minterms and noise products:

$\Sigma_N = N^1_{x_1}N^2_{\overline{x_1}}$

The valid minterm hyperspace is as follows:

$\tau_N = N^1_{x_1}N^2_{x_1} + N^1_{\overline{x_1}}N^2_{\overline{x_1}}$

The noise vectors in $\Sigma_N$ and $\tau_N$ are orthogonal as they do not contain any common minterms. The result $S_N = \tau_N \cdot \Sigma_N$ will have a zero average, concluding that this example is unsatisfiable.

\begin{theorem} If the product of the NBL-SAT instance $\Sigma_N$ and hyperspace $\tau_N$
  produces a zero average, then $S = f(x_1, x_2, \cdots, x_M)$ is unsatisfiable.
\begin{proof}
If a subset of clauses $\{c_j, c_{j+1}, \cdots, c_k\}$ in $S$ are unsatisfiable, then there are no common minterms among $c_j, c_{j+1}, \cdots, c_k$.
As such, the corresponding noise vectors $Z^j, Z^{j+1}, \cdots, Z^k$, which contain the additive superposition of minterms that satisfy 
$c_j, c_{j+1}, \cdots, c_k$ respectively, will form a superposition of logically invalid noise minterms (i.e. $N^j_{x_i}N^k_{\overline{x_i}}$). As $\tau_N$ contains only valid minterms by construction, $\Sigma_N$ and $\tau_N$ will be 
uncorrelated, and the product $\tau_N \cdot \Sigma_N$ will produce a zero average output.
\end{proof}
\label{theorem:unsat}
\end{theorem}

Note that two key observations can be made at this stage:

\begin{itemize}
\item Applying the test of Theorem~\ref{theorem:unsat} allows us to determine if
  $S$ is SAT with a single operation. 
\item The reason why we are able to perform the SAT check with a single operation is that we are
  able to effectively and simultaneously apply all minterms to the NBL-SAT instance,
  since each of the minterms in NBL are orthogonal basis noise
  vectors. This is not possible in traditional SAT solvers. 
\end{itemize}

\subsection{Algorithm to Determine Satisfying Assignment using NBL-SAT}
\label{sec:sat-assignment}

Algorithm~\ref{algo:assign} describes the NBL-SAT procedure to determine the
satisfying assignment for a SAT instance $S$. It is assumed that
Algorithm~\ref{algo:check} has been run and has shown $S$ to be satisfiable before Algorithm~\ref{algo:assign}
is invoked.

\begin{algorithm}[htb]
 \begin{small}
    \begin{algorithmic}[1]
      \STATE{$NBL-SAT\_satisfying\_assignment\_determination(S_N)$}
      \STATE{$Result = \phi$}
      \FOR{$i$ = 1 to $n$}
      \STATE{$\tau_N^{red} \leftarrow (\tau_N$ with variable $x_i$ bound to 1)}
      \STATE{$S_N^{red} \leftarrow (\tau_N^{red} \cdot \Sigma_N$)}
      \IF{NBL-SAT\_check($S_N^{red}$) is unsatisfiable}
      \STATE{$Result \leftarrow Result \cup \overline{x_i}$}
      \STATE{$\tau_N^{red} \leftarrow (\tau_N$ with variable $x_i$ bound to 0)}
      \ELSE
      \STATE{$Result \leftarrow Result \cup x_i$}
      \ENDIF
      \STATE{$\tau_N \leftarrow \tau_N^{red}$}
      \ENDFOR
      \STATE{return $Result$}
    \end{algorithmic}
    \caption{Pseudocode of NBL-SAT satisfying assignment determination}
    \label{algo:assign}
 \end{small}
\end{algorithm}

Algorithm~\ref{algo:assign} starts by initializing the result to $\phi$
(line 2). We iterate over all $n$ variables of
the problem (line 3). In the $i^{th}$ iteration, we take the current
reduced hyperspace $\tau_N$, and  bind the variable $x_i$
to 1 (line 4). By binding the
variable $x_i$ to 1, we limit the reduced hyperspace $\tau_N^{red}$ to contain only valid minterms in the $x_i$ subspace.
In essence, we are 
testing to see whether the current $S_N^{red}$ has a satisfying solution in the $x_i$ subspace. 
If the NBL-SAT\_check of $S_N^{red}$ returns "unsatisfiable", then the
solution is in the $\overline{x_i}$ subspace (since $S_N$ is known to be
satisfiable a-priori, given that that Algorithm~\ref{algo:check} has already
been run). Hence we append $\overline{x_i}$  to the result (line 7), and continue
further processing after binding variable $x_i$ to 0 (line 8). If $\Sigma_N^{red}$ is satisfiable,
then the solution is in the $x_i$ subspace, and we append $x_i$ to the
result (line 10). Before continuing the next iteration, we update $\tau_N$
with $\tau_N^{red}$ (line 12), to ensure that future iterations inherit the
variable binding that was conducted in the current iteration. The result is
finally returned in line 14.

{\bf Example 8:} Consider the SAT instance of Example 6,  $S = (x_1 + \overline{x_2}) \cdot (\overline{x_1} + \overline{x_2})$, which has been
known to be satisfiable according to Algorithm~\ref{algo:check}. The NBL-SAT formula is as follows:

$\Sigma_N = (N^1_{x_1}N^1_{x_2} + N^1_{x_1}N^1_{\overline{x_2}} + N^1_{\overline{x_1}}N^1_{\overline{x_2}}) \cdot (N^2_{x_1}N^2_{\overline{x_2}} + N^2_{\overline{x_1}}N^2_{x_2} + N^2_{\overline{x_1}}N^2_{\overline{x_2}})$

$\tau_N = (N^1_{x_1}N^2_{x_1} + N^1_{\overline{x_1}}N^2_{\overline{x_1}}) \cdot (N^1_{x_2}N^2_{x_2} + N^1_{\overline{x_2}}N^2_{\overline{x_2}})$

$S_N = \Sigma_N \cdot \tau_N$

$\Sigma_N$ contains two valid noise minterms $N^1_{x_1}N^2_{x_1}N^1_{\overline{x_2}}N^2_{\overline{x_2}}$, $N^1_{\overline{x_1}}N^2_{\overline{x_1}}N^1_{\overline{x_2}}N^2_{\overline{x_2}}$ which are $x_1\,\overline{x_2}, \overline{x_1}\,\overline{x_2}$ respectively.

Now in the first iteration of Algorithm~\ref{algo:assign}, we bind variable $x_1$ to 1 (line 4), yielding 

$\tau_N^{red} = (N^1_{x_1}N^2_{x_1} + 0) \cdot (N^1_{x_2}N^2_{x_2} + N^1_{\overline{x_2}}N^2_{\overline{x_2}})$

$\tau_N^{red} = N^1_{x_1}N^2_{x_1}N^1_{x_2}N^2_{x_2} 
+ N^1_{x_1}N^2_{x_1}N^1_{\overline{x_2}}N^2_{\overline{x_2}}$

$S_N^{red} = \Sigma_N \cdot \tau_N^{red}$

Thus $S_N^{red}$ has a positive average value as the noise minterm $N^1_{x_1}N^2_{x_1}N^1_{\overline{x_2}}N^2_{\overline{x_2}}$ exists in both $\Sigma_N$ and $\tau_N^{red}$. The NBL-SAT\_check will return that $S_N^{red}$ is satisfiable, and $x_1$ is appended to the (initially empty) result (line 7) and we update $\tau_N \leftarrow \tau_N^{red}$.

In the second iteration, we bind variable $x_2$ to 1, yielding

$\tau_N^{red} = (N^1_{x_1}N^2_{x_1} + 0) \cdot (N^1_{x_2}N^2_{x_2} + 0)$

$\tau_N^{red} = N^1_{x_1}N^2_{x_1}N^1_{x_2}N^2_{x_2}$

$S_N^{red} = \Sigma_N \cdot \tau_N^{red}$

The $S_N^{red}$ has a zero average value and the NBL-SAT\_check will return that $S_N^{red}$ is unsatisfiable, hence $\overline{x_2}$ is appended
to the final result $x_1\,\overline{x_2}$ (line 10) which is our satisfying assignment for the example.

Note that Algorithm~\ref{algo:assign} yields a satisfying minterm. It can
be easily modified to return satisfying cubes. To do this, each iteration
would bind a variable $x_i$ to both 1 and 0. If the resulting
$S_N^{red}$ outputs both have a positive average value, then variable $x_i$ would be omitted from the result.

\subsection{Scaling Issues}

In order to discuss how NBL-SAT scales with the number of variables and
clauses, consider  3-SAT instances (in which each clause has 3 literals)
with $n$ variables and $m$ clauses. We assume that each basis noise source
($N^j_{x_i}$) is a uniform random variable between [-0.5, 0.5]. Recall that
the average value of $\tau_N \cdot \Sigma_N$ is proportional to the number
of satisfying minterms, since such minterms are present in both $\tau_N$
and $\Sigma_N$. Hence the ability of NBL-SAT to discriminate between an
instance with one satisfying minterm and another instance which is
unsatisfiable needs to be considered. We define the SNR of NBL-SAT as 

SNR = $\frac{\hat{\mu_1} - 3 \hat{\sigma_1}}{\hat{\mu_0} + 3 \hat{\sigma_0}}$

where $\hat{\mu_i}$ is the expectation of the mean of the average value of $\tau_N \cdot
\Sigma_N$ when there are $i$ satisfying minterms, and $\hat{\sigma_i}$ is the
expectation of the standard deviation of the average value of $\tau_N \cdot
\Sigma_N$, when there are $i$ satisfying minterms. Note that $\hat{\mu_0}$ = 0.
Assuming that there are $N$ samples in each noise source, we have:

$\hat{\mu_1} = E(\frac{1}{N} \Sigma_{i=1}^N \{ \Pi_{j = 1}^{nm} x_j^2 \} )$

where $x_j$ is uniformly distributed within [-0.5, 0.5]. The product is
over $nm$ since there are $nm$ noise products in any satisfying minterm in
NBL-SAT. Simplifying, we have $\hat{\mu_1} = (\frac{1}{12})^{nm}$. 

Similarly, the unbiased estimate of the variance of the mean of the product
of $nm$
independent uniform distributions~\cite{Dettmann2009} (over $N$ samples) is given by $\hat{\sigma^2} = \frac{1}{N-1}
(\frac{1}{12})^{2nm}$.

Now the total number of products in a NBL-based 3-SAT instance with $n$ variables
and $m$ clauses is $(2^n)\cdot (2^n - 2^{n-3})^m$ $\sim$ $O(2^{nm})$. The first term
refers to the number of products of $\tau_N$, while the second term is the
number of products in $\Sigma_N$. Since these $O(2^{nm})$ products are
independent, their variances will add up, and so we have $\hat{\sigma_1} = \hat{\sigma_0} = \frac{1}{\sqrt{N-1}}
(\frac{1}{12})^{nm} \cdot 2^{nm}$.

For SNR $\gg$ 1, we can ignore $\hat{\sigma_1}$ in the SNR expression, yielding

SNR = $\frac{\hat{\mu_1}}{3 \hat{\sigma_0}} = \frac{\sqrt{N-1}}{3 \cdot 2^{nm}}$.

Note that if it is known that the instance has $K$ satisfying minterms,
then the SNR expression above is multiplied by $K$.


%% file: experiment.tex
\section{Experimental Results}
\label{sec:experiment}

To validate our NBL-SAT algorithm, we simulated several small NBL-SAT $S_N$ instances and corresponding $T_N$ hyperspaces in MATLAB.
In our simulations, each basis noise source ($N^j_{x_i}$) is a uniform random variable 
between [-0.5, 0.5]. Each instance is simulated until the mean value of $S_N$ has converged to the 
third significant digit or until $10^8$ noise samples have been reached. Our experiments focus
on the SAT checker from Algorithm 1, as the satisfying assignment determination from Algorithm 2
simply consists of iterative applications of the SAT checker.

We use the following two examples, one unsatisfiable and one satisfiable, to validate the correctness of our scheme.

$S_{UNSAT} = (x_1 + x_2) \cdot (x_1 + \overline{x_2}) \cdot (\overline{x_1} + x_2) \cdot (\overline{x_1} + \overline{x_2})$

$S_{SAT} = (x_1 + \overline{x_2}) \cdot (\overline{x_1} + \overline{x_2}) \cdot (x_1 + \overline{x_2}) \cdot (\overline{x_1} + \overline{x_2})$ 

\begin{figure}
\begin{center}
\resizebox{1.0\columnwidth}{!}{\epsfig{figure=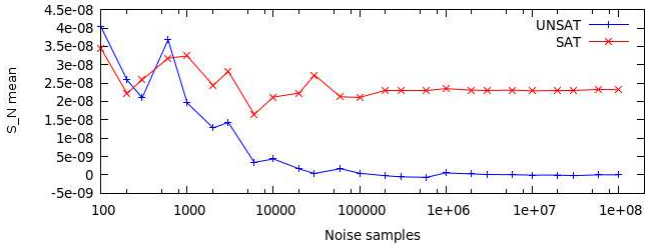}}
\vspace{-0.3in}
\caption{$S_N$ mean for UNSAT and SAT instances}
\label{fig:sn_plot}
\end{center}
\vspace{-0.3in}
\end{figure}

The first clause in our satisfiable example is redundant, but brings the number of clauses $m$ to 4 and make the $S_N$ values comparable with our unsatisfiable example which also has $m=4$. In Figure~\ref{fig:sn_plot}, the average mean values of $S_N$ of both examples are plotted as a function of number of noise samples.

%% file: discussion.tex
\section{Realizing an NBL-based SAT Engine}
\label{sec:discussion}

The NBL-based SAT algorithm is easily realized using existing hardware and
software based approaches. We devote this section to a discussion on
possible realizations. 

A first observation we make in this regard is that instead of using
uncorrelated noise sources as the basis vectors, we could utilize
sinusoidal signals as the basis vectors~\cite{lkthn2, Bollapalli2010}. Assuming that the highest
frequency sinusoid realizable in today's technology has a frequency $F$,
(typically in the 10s of GHz), and that all the basis sinusoids are
equi-spaced with a frequency difference of $f$ between adjacent sinusoids,
we could realize $F/f$ variables for the Sinusoid-based Logic (SBL) SAT
engine. Minimizing $f$ would be a key design criterion, allowing us to
implement a large number of variables. A small value of $f$ would require
the low-pass filters of high order, yielding a more complex circuit.
The tradeoff of circuit complexity versus number of variables remains
an open exercise.

Using the above ideas, a hardware based NBL-based SAT field-programmable
engine can be envisioned as well. Such an engine would have a plurality of adders
(implementing configurable clauses), multipliers (implementing the
conjunction operation among the clauses), and noise sources (which could
potentially be made up of wideband amplifiers which amplify a resistor's
thermal noise, or realized using pseudorandom number generator). In an SBL based engine, on-chip sinusoidal oscillators of different
frequencies~\cite{Mandal2011, vc-wave, vinay-ppr} could be utilized. Such an
engine  would have an on-chip 
correlator block as well. Having such a reconfigurable engine would
allow the user to load their specific SAT instance on this engine, and run
it using the algorithms described in Section~\ref{sec:approach}.

A natural extension to the hardware based NBL-based SAT engine is a
hybrid approach using both CPU and a NBL-based SAT coprocessor as in~\cite{Gulati2008, Gulati2010},
where in the primary (exact) SAT solver is implemented on the CPU, but the assignment of variables is guided 
through the NBL-SAT coprocessor. For example, we could iterate over all variables where each variable is bound to 1 and 0, 
and check in the NBL-SAT coprocessor if the reduced $S_N$ is satisfiable. As the $S_N$ mean is
directly proportional to the number of satisfying minterms, we choose the binding that 
results in the highest $S_N$ mean, thus potentially improving the efficiency of the CPU SAT solver to 
find an assignment.

Both the NBL-based SAT engine and the SBL based SAT engine could be
simulated in software. In both cases, the addition, multiplication and
autocorrelation operations would be performed in software. The adders,
multipliers and correlators could be implemented as analog blocks or
digital blocks in such a simulation based approach. 

%% file: conclusion.tex
\section{Conclusions}
\label{sec:conclusion}

Noise-based Logic (NBL) is a recently proposed approach to realize logic
circuits. NBL is actually a {\em deterministic} logic system, contrary to what its name may suggest. Among
its most powerful features is the ability to apply all 
$2^n$ input minterms to a $n$-input circuit, {\em simultaneously}. Using NBL, we
have presented a novel approach to solve the Boolean 
Satisfiability (SAT) problem. 
By exploiting the superposition and correlation properties of
noise basis sources in NBL, our approach circumvents a key assumption (and restriction) in the
traditional approach to solving SAT. 
In our NBL-based SAT approach, we show that the decision about whether an instance is SAT or not can
be made in a single operation, and a satisfying solution can be found in linear
number of such operations. 
A key advantage of NBL-SAT algorithm is that an NBL-based SAT engine can be easily implemented using existing hardware and software.
This paper also discusses the scalability of NBL-SAT, and for NBL in general.
Additionally, NBL-SAT is not limited to noise and can be realized using sinusoidal signals, pulse-based logic, and RTW-based logic.
Although no NBL style circuits have been
developed to date, we hope that the power of NBL will encourage work in the realization of
such circuits.